\documentclass[12pt]{iopart}

\usepackage{graphicx}
\usepackage{bm}
\newcommand{\aw}{8cm}
\newcommand{\aww}{9cm}

\def\tr{\,{\rm tr}\,}
\def\ket#1{|#1\rangle}
\def\bra#1{\langle#1|}

\def\vec#1{\bm{#1}}

\begin{document}

\title{Entanglement of random vectors}
\author{Marko \v Znidari\v c}
\address{Physics Department, Faculty of Mathematics and Physics, University of Ljubljana, Slovenia}

\begin{abstract}
We analytically calculate the average value of $i$-th largest Schmidt coefficient for random pure quantum states. Schmidt coefficients, i.e., eigenvalues of the reduced density matrix, are expressed in the limit of large Hilbert space size and for arbitrary bipartite splitting as an implicit function of index $i$.
\end{abstract}

\submitto{\JPA}


Entanglement is a valuable quantum resource and as such a lot of attention has been devoted to studies of entanglement and ways of exploiting entanglement present in various quantum states. Random quantum states are interesting for several reasons. They are used in various quantum information processes like e.g. in quantum communication. Also, a generic quantum evolution falling under the label of quantum chaos will produce a random state from almost arbitrary initial state. It is therefore important to understand entanglement properties of random quantum states. It comes as no surprise that the subject has a long history, for instance, entropy of a subsystem has been calculated in~\cite{Lubkin:78,Lloyd:88,Page:93,Ruiz:95,Sen:96} while purity has been studied in~\cite{Lubkin:78,Zyczkowski:01}. From a practical point of view a relevant question is how to generate random pure states? Several methods have been proposed, using the so-called pseudo-random unitary operators~\cite{Emerson:03,Weinstein:05,Emerson:05,Oliveira:06} which are characterized by the number of random parameters scaling polynomially with the number of qubits and not exponentially as it would be required in order to generate true random unitary operators. Of course, using less than a full set of exponentially many parameters causes pseudo-random operators to be just an approximation to the true ones. Nevertheless, if a common criterion of entanglement such as purity or von Neumann entropy of a subsystem is used, the convergence to the asymptotic values of purity or entropy is observed already after polynomial number of steps. While purity might be a perfectly good benchmark for certain purposes we must nevertheless keep in mind that purity or entropy of a subsystem is just one of many possible entanglement measures of pure states. In fact, to fully characterize entanglement properties of a given bipartite state a full list of Schmidt coefficients is needed. If we have a bipartite Hilbert space ${\cal H}={\cal H}_{\rm A}\otimes {\cal H}_{\rm B}$ of dimension $NK$, with $N$ being dimension of ${\cal H}_{\rm A}$ and $K$ of ${\cal H}_{\rm B}$, the Schmidt decomposition is
\begin{equation}
\ket{\psi}=\sum_{i=0}^{N-1} \sqrt{\lambda_i} \ket{w^{\rm A}_i} \otimes \ket{w^{\rm B}_i},
\label{eq:Schmidt}
\end{equation}    
where $\ket{w^{\rm A}_i}$ and $\ket{w^{\rm B}_i}$ are orthonormal on respective subspaces ${\cal H}_{\rm A}$ and ${\cal H}_{\rm B}$. Squares of Schmidt coefficients $\lambda_i$ are nothing but the eigenvalues of the reduced density matrix, $\rho=\tr_K{\ket{\psi}\bra{\psi}}$, obtained by tracing over subspace ${\cal H}_{\rm B}$. Without sacrificing generality we will assume $N\le K$ as well as that the eigenvalues $\lambda_i, i=0,\ldots,N-1$ of $\rho$ are listed in decreasing order. Schmidt coefficients are invariant under local transformation (transformations acting separably on ${\cal H}_{\rm A}$ and ${\cal H}_{\rm B}$) and fully specify bipartite entanglement of state $\ket{\psi}$. Therefore, a set of $N-1$ entanglement monotones, for instance, sums of $k$ smallest eigenvalues, is needed to characterize pure state entanglement~\cite{Vidal:00}. A sum of $k$ smallest eigenvalues is also a good estimate for the truncation error in recently proposed numerical method~\cite{Vidal:03} for simulation of weakly entangled quantum systems using matrix product state representation~\cite{MPS}. For random states joint distribution $P_{\rm N,K}(\lambda_0,\ldots,\lambda_{N-1})=P_{\rm N,K}(\vec{\lambda})$ of eigenvalues is a well known quantity~\cite{Lloyd:88,Zyczkowski:01}, see equation (\ref{eq:P}) below. However, it is not particularly convenient to deal directly with the distribution in $N$ dimensional space. It would be much simpler to have some scalar quantities which could be used as a benchmark for the randomness of quantum states. Because the eigenvalues fully determine entanglement we are going to calculate the average value of $\lambda_i$ for random quantum states, i.e., the average value of $i$-th largest eigenvalue of the reduced density matrix, $\langle \lambda_i \rangle$, where the angle brackets denote averaging over random states. Random vectors $\ket{\psi}$ in ${\cal H}$ can be obtained in two equivalent ways. As columns of random unitary matrices distributed according to the invariant Haar measure, which can in turn be obtained by, e.g., Hurwitz parametrization of the unitary group. A second, simpler procedure, is to simply draw real and imaginary parts of expansion coefficients of $\ket{
\psi}$ as random Gaussian variables, subsequently normalizing the vector~\cite{Zyczkowski:01}. We will be especially interested in the limit of large dimensions $N$. Because $\lambda_i$ scale as $\sim 1/N$ (due to $\sum_i \lambda_i=1$), we will define a function $f(x)$, such that
\begin{equation}
\lambda(x) =\langle \lambda_i \rangle = \frac{1}{N} f(x), \qquad x=\frac{i+1/2}{N}.
\label{eq:fdef}
\end{equation}  
In the limit $N \to \infty$ the function $f(x)$ becomes a continuous function with the definition range $x\in[0,1]$\footnote{The reason to set $x=(i+1/2)/N$ and not e.g. $x=i/N$ is purely empirical. By comparing our analytical solution for $f(x)$ with numerics for finite $N$ we have found that setting $x=(i+1/2)/N$ gives better agreement than e.g. $x=i/N$ or $x=(i+1)/N$. Note though, that the difference between all these choices of $x$ goes to zero with $N \to \infty$.}. Deriving and studying the function $f(x)$ is the main aim of this paper. There are several ways of calculating $\langle \lambda_i \rangle$. The most direct one is by using joint density of eigenvalues $P_{N,K}(\vec{\lambda})$. While such method is exact for all $N$ it is rather complicated. As we are mainly interested in the asymptotic $N \to \infty$ behaviour we are going to use simpler method to derive $f(x)$ in the second part of the paper.
\par
For random $\ket{\psi}$ the distribution of eigenvalues is~\cite{Lloyd:88,Zyczkowski:01}
\begin{equation}
\fl P_{N,K}(\vec{\lambda})=\frac{\Gamma{(NK)}}{\prod_{i=0}^{N-1} \Gamma(K-j)\Gamma(N-j+1)}\delta(1-\sum_{i=0}^{N-1}\lambda_i)\prod_i \lambda_i^{K-N} \prod_{i<j} (\lambda_i-\lambda_j)^2.
\label{eq:P}
\end{equation}
To calculate the average of $i-$th largest eigenvalue $\lambda_i$ or its distribution it is useful to first define the probability $E(i;s)$ that exactly $i$ eigenvalues lie in the interval $[0,s]$. Then the probability density of $i$-th eigenvalue is given by $F(i;s)$, obtained as
\begin{equation}
F(0;s)=\frac{{\rm d}E(0;s)}{{\rm d}s},\qquad F(i;s)-F(i-1;s)=\frac{{\rm d}E(i;s)}{{\rm d}s}, \hbox{ for }i>0.
\end{equation}
For standard Gaussian random matrix ensembles the distribution of the largest eigenvalue has been calculated in~\cite{Tracy:94,Tracy:96}. Average value of $i$-th eigenvalue is then simply $\langle \lambda_i \rangle=\int\! s F(i;s) {\rm d}s$. For instance, the smallest and the largest eigenvalues are given by
\begin{eqnarray}
\langle \lambda_{\rm min} \rangle &=& N \int_0^1{\!\!{\rm d}s\, s \int_s^1{P_{N,K}(s,\vec{\lambda})}{\rm d}\vec{\lambda}} \nonumber \\
\langle \lambda_{\rm max} \rangle &=& N \int_0^1{\!\!{\rm d}s\, s \int_0^s{P_{N,K}(s,\vec{\lambda})}{\rm d}\vec{\lambda}},
\label{eq:int}
\end{eqnarray}
where the inner integral is over $N-1$ arguments of $P_{N,K}$ (\ref{eq:P}). Using these formulae in the simplest case of a symmetric bipartite cut with $N=K$ one obtains for $N=2$ $\langle \lambda_0 \rangle=\frac{7}{8}$ and $\langle \lambda_1 \rangle=\frac{1}{8}$, while for $N=3$ one gets $\langle \lambda_0 \rangle=\frac{313}{432}$, $\langle \lambda_1 \rangle=\frac{103}{432}$ and $\langle \lambda_2 \rangle=\frac{1}{27}$. We have calculated $\langle \lambda_{\rm min} \rangle$ for several other small values of $N$ with the result always being $1/N^3$. We therefore conjecture that for $N=K$ we have in general 
\begin{equation}
\langle \lambda_{\rm min} \rangle = \frac{1}{N^3}.
\label{eq:min}
\end{equation}
Even though the integrals in (\ref{eq:int}) can be connected to Laguerre polynomials~\cite{Mehta:91,Zyczkowski:01} they are non-trivial because the inner integrations are over a finite range. To calculate $f(x)$ in the limit $N \to \infty$ we will use a different approach. Instead of using the exact $P_{N,K}$ and afterwards taking the limit we will instead use the density of eigenvalues $p(\lambda)$ which is already evaluated in the $N,K \to \infty$ limit.
\par
Eigenvalue density $p(\tau)$ of scaled eigenvalues $\tau=\lambda_i N$ can be obtained by integrating out $N-1$ eigenvalues in $P_{\rm N,K}$. For large $N$ and $K$ the result is~\cite{Page:93}
\begin{equation}
p(\tau)=\frac{\sqrt{(\tau-a)(b-\tau)}}{2\pi \tau},
\label{eq:p}
\end{equation}
if $\tau \in [a,b]$, otherwise $p(\tau)=0$. Parameters $a,b$ are given as
\begin{equation}
a=(1-\sqrt{w})^2,\qquad b=(1+\sqrt{w})^2,\qquad w=\frac{N}{K}.
\label{eq:w}
\end{equation}
The above density of eigenvalues (\ref{eq:p}) is well known in statistics, where it occurs as eigenvalue density of random covariance matrices $W W^\dagger$, also known as Wishart matrices, and is called Mar\v cenko-Pastur law~\cite{Marcenko:67}. In the simplest case of an equal partition, $N=K$, we have $a=0$ and $b=4$, resulting in $p(\tau)=\frac{1}{2\pi}\sqrt{4/\tau-1}$. Moments of eigenvalue distribution, $\langle \lambda^\nu \rangle=\tr{\rho^\nu}$ have been calculated in~\cite{Sommers:04}. The calculation of $f(x)$ from $p(\tau)$ is simple because $f(x)$ is by definition a decreasing function of $x$. Noting that $p(\tau) = 1/|f'(x^*)|$, where $x^*$ is such that $f(x^*)=\tau$, we can immediately write differential equation for $f(x)$,
\begin{equation}
f'(x)=-1/p(x),
\label{eq:f'}
\end{equation}
with $f'(x)={\rm d}f(x)/{\rm d}x$ and the boundary conditions $f(0)=b$ and $f(1)=a$. Let us first study the symmetric case of equal bipartition, $N=K$, and then separately the general $N<K$.
\par
\begin{figure}[h]
\centerline{\includegraphics[width=\aww]{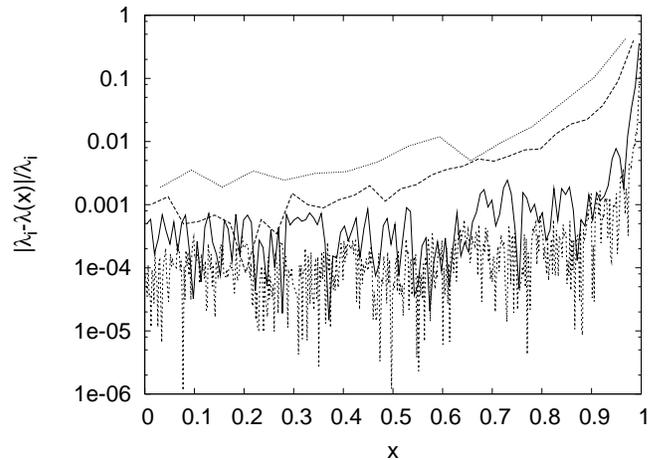}}
\caption{Relative error of numerically calculated average $\lambda_i$ for $N=2^n$, $n=4,5,7,9$ (top to bottom), as compared to the theoretical formula for $f(x)$ (\ref{eq:f}). Relative error decreases as $1/N$.}
\label{fig:ran_err}
\end{figure}

Integrating (\ref{eq:f'}) once, using (\ref{eq:p}) and setting $N=K$, we get $f(x)$ in an implicit form
\begin{equation}
f(x)=4 \cos^2{\varphi},\qquad \frac{\pi}{2}x=\varphi-\frac{1}{2}\sin{(2\varphi)},\qquad \{N=K\}.
\label{eq:f}
\end{equation}
\begin{figure}[h]
\centerline{\includegraphics[width=\aw]{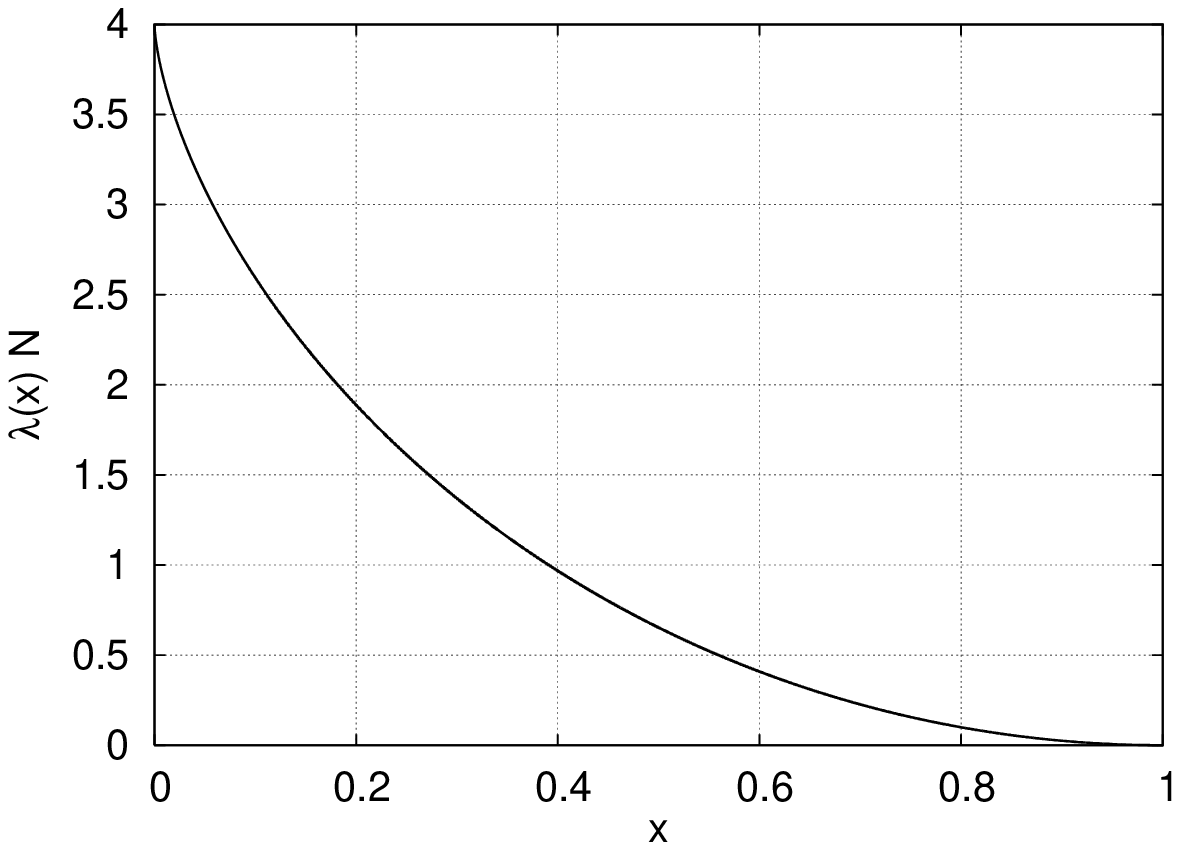}\includegraphics[width=\aw]{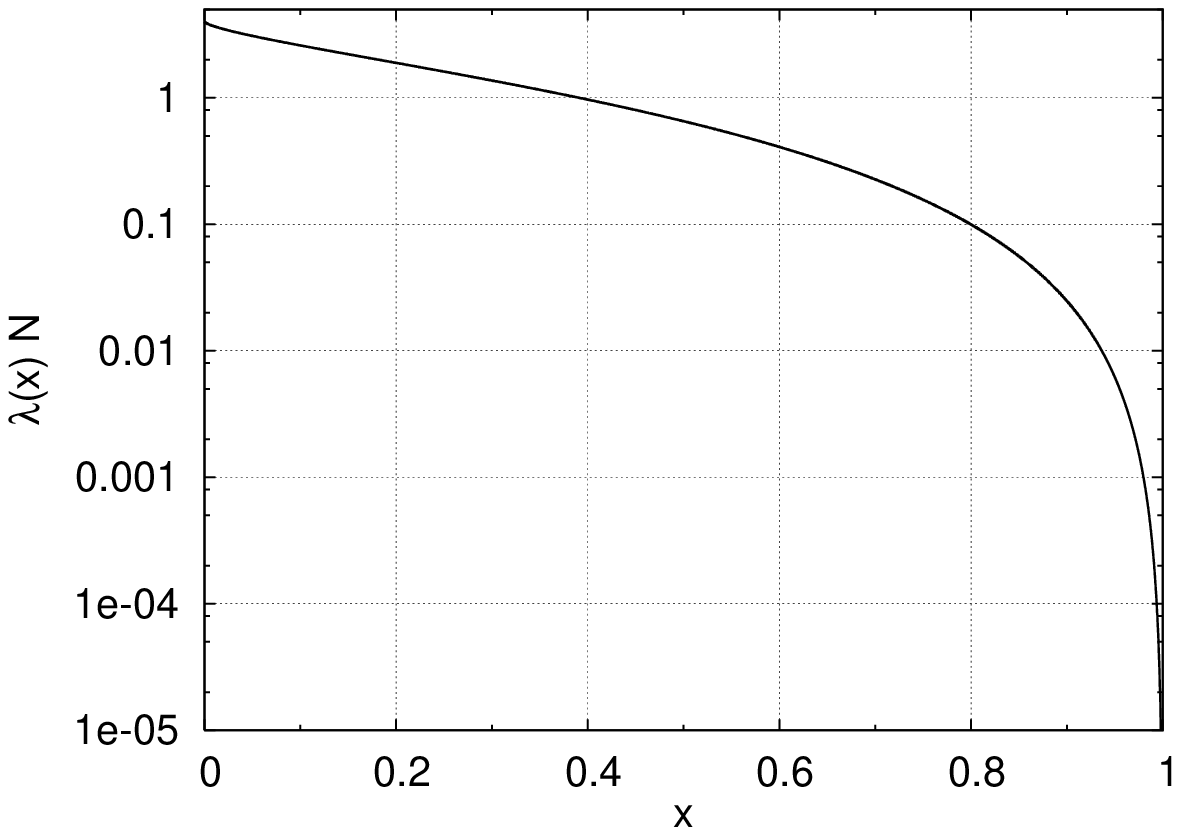}}
\caption{Theoretical $f(x)=N\lambda(x)$ (\ref{eq:f}) for $N=K$. In the right frame is a semi-log plot.}
\label{fig:ran_lam}
\end{figure}
Functional form of $f(x)$ (\ref{eq:f}) can be seen in figure~\ref{fig:ran_lam}. For small or for large values of $x$, i.e., for large and small eigenvalues, one can expand implicit solution (\ref{eq:f}) to get $f(x)\approx4-4(3\pi x/4)^{3/4}$ for $x\ll 1$ and $f(x)\approx \pi^2(1-x)^2/4$ for $1-x \ll 1$. We see that the largest eigenvalue is $\langle \lambda_0 \rangle \sim 4/N$, while the smallest one scales as $\langle \lambda_{N-1}\rangle\sim 1/N^3$, in accordance with the conjecture (\ref{eq:min}). Theoretical formula for $f(x)$ (\ref{eq:f}) gives asymptotic $N \to \infty$ result. What about the eigenvalues for finite $N$? We performed numerical calculation of average $\lambda_i$ by randomly drawing vectors and then calculating the eigenvalues of the reduced density matrix. The convergence to the asymptotic $f(x)$ is pretty fast and already for moderate $N=2^n$ with $n=4$ ($NK=2^8$), theoretical formula (\ref{eq:f}) agrees with numerical $\lambda_i$ to better than 1\%. In figure~\ref{fig:ran_err} we can see that the relative error decreases as $\sim 1/N$ apart from the smallest eigenvalue (and possibly few smallest ones) for which theoretical formula (\ref{eq:f}) gives $\lambda_{N-1} \approx 0.62/N^3$ while the exact value is conjectured to be $1/N^3$ (\ref{eq:min}). 
\begin{figure}[h]
\centerline{\includegraphics[width=\aww]{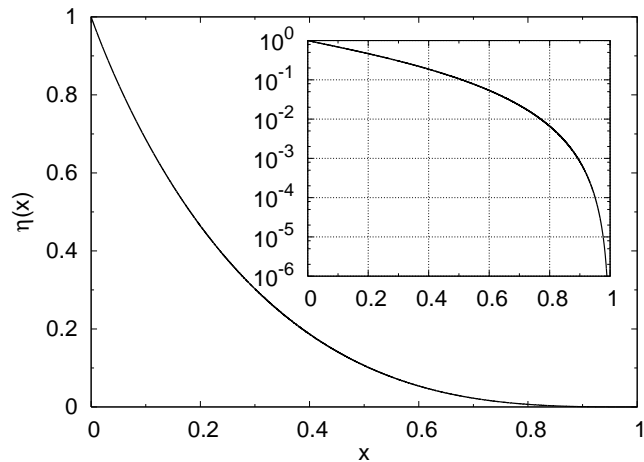}}
\caption{Theoretical error $\eta(x)$ (\ref{eq:eta},\ref{eq:f}) committed by retaining a fraction $x$ of the largest eigenvalues. In the inset is a semi-log plot of the same $\eta(x)$. All for $N=K$.}
\label{fig:eta}
\end{figure}
We have also numerically calculated the width $\delta\lambda_i$ of the distribution of $\lambda_i$, $(\delta \lambda_i)=\sqrt{\langle \lambda_i^2 \rangle - \langle \lambda_i \rangle^2}$. Numerical data (not shown) indicate that the width $\delta \lambda_i$ is approximately equal to one half of the spacing $\lambda_i-\lambda_{i-1}$. Using theoretical formula for the spacing (i.e. for $f'(x)$) we have found that an approximate relation $\delta \lambda_i \approx 4/(N^2\sqrt{4/f(x)-1})$ holds, where $f(x)$ is given in (\ref{eq:f}). With increasing $N$ the relative width decreases and therefore the eigenvalues become increasingly peaked around their average $\langle \lambda_i \rangle$. If we want to approximate a given state with less than a full set of parameters, e.g. as a matrix product state~\cite{MPS}, the quantity that estimates an error made by retaining only a fraction $x$ of largest eigenvalues is $\eta(x)$, defined as
\begin{equation}
\eta(x)=\int_x^1{\!\!{\rm d}x\, \lambda(x)}.
\label{eq:etadef}
\end{equation}    
Changing integrating variable from $x$ to $\varphi$ the above integral can be evaluated. Using our $f(x)$ we get
\begin{equation}
\eta(x)=1-\frac{1}{\pi}\{ 2\varphi-\frac{1}{2}\sin{(4\varphi)}\},
\label{eq:eta}
\end{equation}
with $\varphi$ being the solution of equation (\ref{eq:f}). Plot of $\eta(x)$ (\ref{eq:eta}) is shown in figure~\ref{fig:eta}. An interesting thing we can see is that because $f(x)$ behaves as $\sim (x-1)^2$ close to $x=1$, the smallest eigenvalues contribute very little to $\eta(x)$. For instance, the first 80\% of eigenvalues sum to more than $0.993$, i.e., discarding 20\% of the smallest eigenvalues will cause fidelity error of no more than $1-F \approx 0.007$. 
\par
Let us now go to the general case of $N\le K$. By a similar procedure as for symmetric $N=K$ situation we arrive at the implicit formula for $f(x)$,
\begin{eqnarray}
 f(x)&=&a+(b-a)\cos^2{\varphi},\qquad \{N\le K \} \nonumber \\
\frac{\pi}{2}x&=&\frac{1+w}{2w}\varphi-\frac{1}{2\sqrt{w}}\sin{(2\varphi)}-\frac{1-w}{2w}\arctan{\left(\sqrt{\frac{a}{b}}\tan{\varphi} \right)},
\label{eq:fg}
\end{eqnarray}
where $a,b$ and $w$ are given in (\ref{eq:w}). By comparing matrix eigenvalue equation for Wishart matrices (i.e., $W W^\dagger$, which is nothing but the unnormalized reduced density matrix in our language) and zeros of Laguerre polynomials in the limit $N,K \to \infty$, it has been shown in~\cite{Dette:01} that the two converge to one another. Therefore, in the limit $N,K \to \infty$ the $i$-th zero of a generalized Laguerre polynomial $L_N^{(K-N+1)}(Kx)$ converges to $N\lambda_i$, i.e., to $f(x_i=(i+1/2)/N)$ with $f(x)$ given in equation (\ref{eq:fg}).

\begin{figure}[h]
\centerline{\includegraphics[width=\aww]{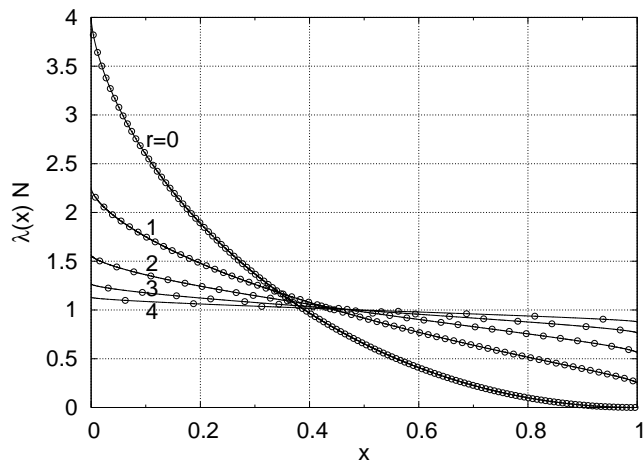}}
\caption{Scaled eigenvalues $\lambda(x)$ for a bipartite cut to $n/2-r$ and $n/2+r$ qubits. Circles are numerics for $n=14$ while the full line is theoretical $f(x)$ (\ref{eq:fg}).}
\label{fig:allr}
\end{figure}
To verify theoretical formula (\ref{eq:fg}) we performed numerical calculation for random vectors in $2^n$ dimensional Hilbert space, making a bipartite cut to $n/2-r$ and $n/2+r$ qubits, i.e., $N=2^{n/2-r}$, $K=2^{n/2+r}$ and $w=1/2^{2r}$. Comparison of numerics and theory for $r=0,1,2,3$ and $4$ is in figure~\ref{fig:allr}. One can see that with increasing $r$, as the subspaces ${\cal H}_{\rm A}$ and ${\cal H}_{\rm B}$ become of increasingly different size, the reduced density matrix on ${\cal H}_{\rm A}$ has increasingly flat spectrum. When one of the subspaces is much larger than the other, that is when $w=N/K \ll 1$, we can simplify the general equation for $f(x)$ (\ref{eq:fg}) to arrive at
\begin{equation}
f(x)\approx 1+2\sqrt{\frac{N}{K}}\cos{(2\varphi)},\qquad \frac{\pi}{2}x=\varphi-\frac{1}{4}\sin{(4\varphi)}\qquad \{N \ll K\}.
\label{eq:fasim}
\end{equation}
For small $w$ the eigenvalues of $\rho$ therefore deviate from $1/N$ by $\sim \frac{2}{N}\sqrt{w}$. The fact that with increasing $r$ all eigenvalues are essentially $\approx 1/N$ is well known. Namely, if we trace a random pure state over large ``environment'' the resulting density matrix will be almost proportional to the identity, e.g., its purity is close to $1/N$~\cite{Lubkin:78,Kendon:02}. In addition, deviations from this totally mixed state go to zero if $K$ increases at a fixed $N$. This is a consequence of the so-called measure concentration in many dimensional spaces~\cite{Hayden:04}~\footnote{For uniform measues in high dimensional space deviations from the average values are very small. As a simple example: probability that the sum of $d$ bounded random numbers deviates by more than $\varepsilon$ from its average is exponentially small in the dimension $d$ as well as in $\varepsilon^2$.}.

\begin{figure}[h]
\centerline{\includegraphics[width=\aw]{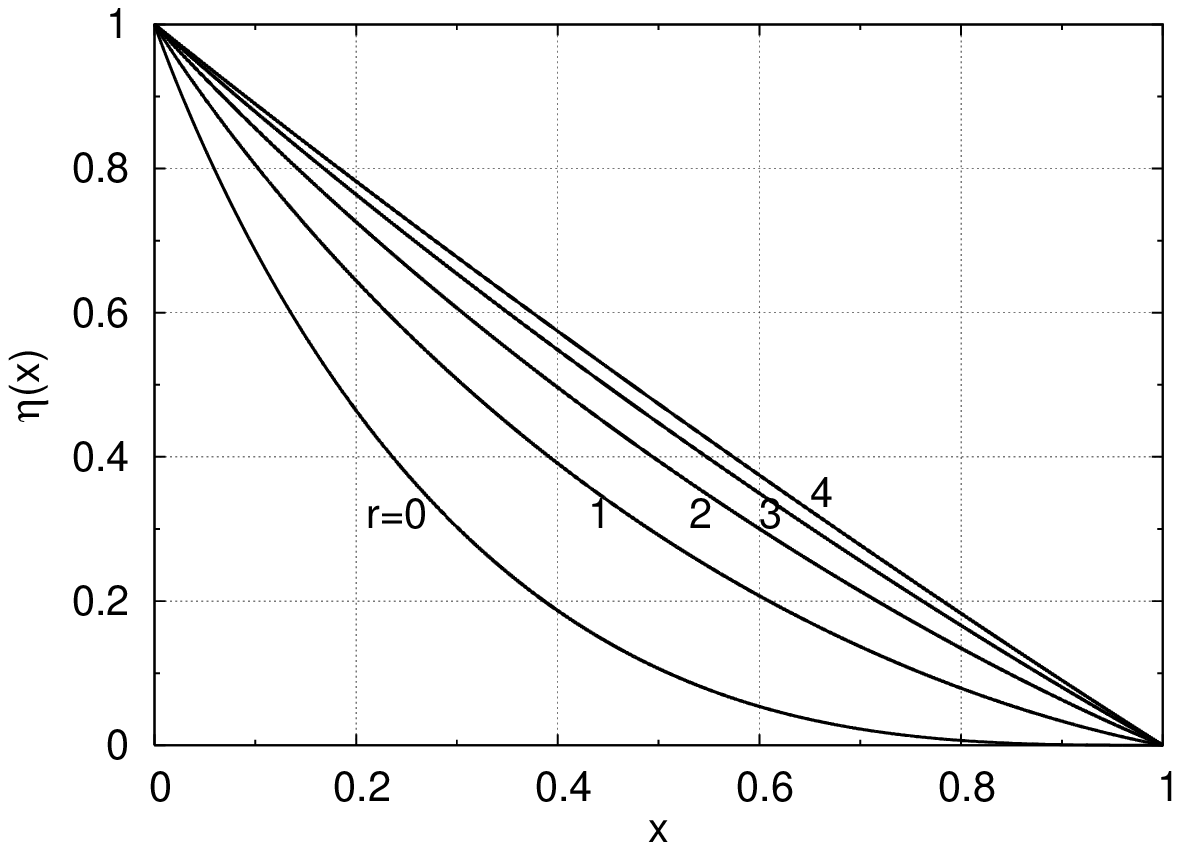}\includegraphics[width=\aw]{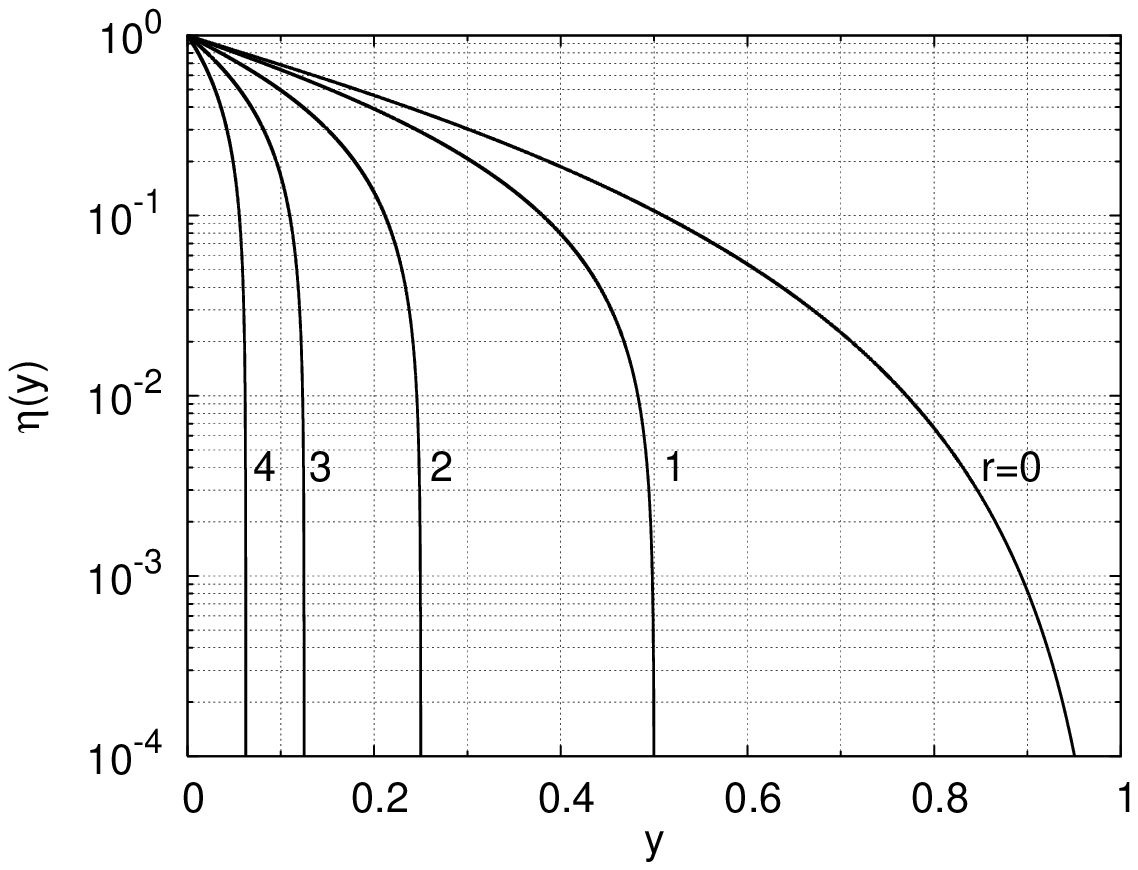}}
\caption{Theoretical error $\eta(x)$ (\ref{eq:eta},\ref{eq:fg}) for the same Hilbert space splitting as in figure~\ref{fig:allr}. In the right frame $\eta$ is plotted as a function of $y=xN/\sqrt{NK}$ so that a given $y$ represents the same number of eigenvalues for all $r$ and not the same fraction, as in the left plot.}
\label{fig:etar}
\end{figure}
The error $\eta(x)$ (\ref{eq:etadef}) made by retaining only $xN$ largest eigenvalues is for $N<K$ given by exactly the same expression as for $N=K$ (\ref{eq:eta}), the only difference is that $\varphi$ is given by the solution of (\ref{eq:fg}). Because the spectrum of $\rho$ becomes flat for large $r$, $\eta(x)$ approaches linear function. This can be seen in figure~\ref{fig:etar}. As the number of retained eigenvalues $xN=x 2^{n/2-r}$ depends on $r$, we also plot in the right frame of figure~\ref{fig:etar} dependence of $\eta$ on the absolute number of retained eigenvalues. From this plot one can see that for instance, keeping $0.2\sqrt{NK}$ largest eigenvalues in all bipartite cuts, we make an error $\eta \approx 0.45$ for symmetric cut with $r=0$, error $\eta \approx 0.40$ for $r=1$, $\eta \approx 0.12$ for $r=2$, while for larger $r$'s the error is zero because the dimension $N$ of ${\cal H}_{\rm A}$ is smaller than the number of retained eigenvalues $0.2\sqrt{NK}$. 
\ack
The author would like to thank Martin Horvat for discussions and Toma\v z Prosen for reading the manuscript as well as Slovenian Research Agency, programme P1-0044, and grant J1-7437, for support.

\end{document}